\documentclass[
reprint,
amsmath,amssymb,
aps,
prx,
]{revtex4-2}

\usepackage{graphicx}
\usepackage{dcolumn}
\usepackage{bm}
\usepackage{float}
\usepackage{gensymb}
\usepackage{siunitx}
\usepackage[utf8]{inputenc}
\usepackage{setspace}
\usepackage{hyperref}
\usepackage[mathlines]{lineno}

\begin{document}

\preprint{\today}
\title{Topological defect coarsening in quenched smectic-C films \\
        analyzed using artificial neural networks}

\author{Ravin A.~Chowdhury}
\author{Adam A.~S.~Green}
\author{Cheol S.~Park}
\author{Joseph E.~Maclennan}
\author{Noel A.~Clark}
\affiliation{%
Department of Physics and Soft Materials Research Center,
University of Colorado, Boulder, Colorado, 80309, USA
}%

\date{\today}

\begin{abstract}

Mechanically quenching a thin film of smectic-C liquid crystal results in the formation of a  dense array of thousands of topological defects in the director field. The subsequent rapid coarsening of the film texture by the mutual annihilation of defects of opposite sign has been captured using high-speed, polarized light video microscopy. The temporal evolution of the texture has been characterized using an object-detection convolutional neural network to determine the defect locations, and a binary classification network customized to evaluate the brush orientation dynamics around the defects in order to determine their topological signs. At early times following the quench, inherent limits on the spatial resolution result in undercounting of the defects and deviations from expected behavior. At intermediate to late times, the observed annihilation dynamics scale in agreement with theoretical predictions and simulations of the $2$D XY model.

\end{abstract}

\maketitle

\section{\label{sec:intro}Introduction}

Topological defects, which are stable disclinations or dislocations in ordered physical systems, are typically formed as a result of spontaneous symmetry-breaking during phase transitions \cite{chaikin_principles_2013, komura_dynamics_1988}.
The formation and evolution of such defects, which have been predicted, and in some cases observed, in such diverse contexts as cosmology \cite{chuang_cosmology_1991} and condensed matter \cite{rao_topological_2001}, is a classical phenomenon that has been studied in many physical systems, including thin magnetic films \cite{kiselev_chiral_2011} and superfluids \cite{owczarek_topological_1991}. Liquid crystals (LCs) are a particularly convenient medium in which to study the behavior of such defects experimentally, with disclinations easily visualized in both the nematic and tilted smectic phases \cite{chaikin_principles_2013}. The structure and dynamics of topological defects in quasi-two-dimensional liquid crystals is broadly reviewed in \cite{harth_topological_2020}.

Fluid smectics are fundamentally interesting because they can be drawn into extremely thin, freely-suspended films of the order of a few molecular layers thick, allowing the study of physics in two dimensions ($2$D) \cite{young_light-scattering_1978,rosenblatt_temperature_1980,pieranski_review_1993}. In the smectic-A (SmA) liquid crystal phase, the long axes of the molecules are oriented, on average, along the layer normal, while in the smectic-C (SmC) phase they are  tilted from the layer normal, breaking the axial symmetry of the SmA phase and introducing topological complexity. The topology of freely-suspended SmC films may be described by projecting the average molecular long axis (the director) onto the plane of the layers, defining a vectorial orientation field called the $c$-director. When these films are viewed in reflection under crossed polarizers, this orientation field typically creates a schlieren texture, with characteristic, cross-like extinction brushes centered on any topological defects \cite{pindak_macroscopic_1980}.

The visual appearance of defects in SmC films depends, in general, on their topological strength, the illumination conditions, and the relative locations and orientations of the other defects \cite{link_defects_2005}. Several experimental studies of SmC defect dynamics have considered films with only a small number of defects \cite{muzny_direct_1992, silvestre_modeling_2009, wachs_dynamics_2014, missaoui_annihilation_2020}. When there are only a few defects in the field of view, and they are well separated, they can either be identified manually or tracked automatically by cross-correlating the images with synthetically generated templates of model defect textures {\cite{silvestre_modeling_2009}.

However, in dense arrays of topological defects, such as those generated in the quenching experiments described here, the orientation fields around the defects produce irregular and complex schlieren textures in polarized light, making detecting and tracking the defects using the previously implemented techniques impractical.
Machine learning has been shown to be a useful tool enabling object detection in images obtained in such diverse areas as solid-state physics\cite{schmidt_recent_2019}, cellular biology \cite{moenDeepLearningCellular2019,greener_guide_2022}, and in protein folding experiments \cite{alquraishi_machine_2021}. We shall demonstrate here that deep learning can be used to solve the seemingly intractable problem of detecting  topological defects in dense, two-dimensional arrays in LC films.

The analysis of coarsening dynamics in LC systems with large numbers of densely-packed topological defects has been found historically to be challenging in both experimental and numerical studies because of the practical difficulty of detecting the defects.
The coarsening dynamics of model $2$D SmC films with high defect densities have been studied extensively using simulations \cite{yurke_coarsening_1993, jang_annihilation_1995, ginzburg_scaling_1995, burlatsky_scaling_1996, rojas_dynamical_1999}. Experimental studies of defect dynamics in thin, quenched SmC films were reported by Muzny \cite{muzny_thesis_1994}, who described the basic phenomenology of quenching, proposed a mechanism for defect generation, and measured the approach dynamics of defect pairs and the decay of defect number with time following the quench.

In more recent experiments \cite{minor_end--end_2020}, high-speed video microscopy was used to capture the textures of mechanically quenched smectic-C films with much better temporal resolution than in Muzny's experiments.
A preliminary analysis of the evolution of the observed arrays of topological defects  using a convolutional neural network (CNN), a type of artificial neural network that is ideal for image analysis, demonstrated the utility of machine learning but also revealed the limitations imposed by training the network using simulated images, discussed further below.

In the present study, we analyzed the same experimental images but using a newer CNN trained on experimental rather than simulated images to determine the locations  of the defects.
The network was able to predict the defect locations starting at earlier times and with a much higher degree of accuracy than before, measurements verified by comparison with manually determined (human-annotated) coordinates. In addition, a binary classification network was trained to distinguish between defects of opposite topological sign, allowing a comprehensive analysis of the coarsening dynamics in these films.
The observed decay of defect density with time was compared with the XY model, with the predictions of a model proposed by Yurke and co-workers \cite{yurke_coarsening_1993}, and with the results of numerical simulations \cite{jang_annihilation_1995}. At early times ($t<0.4$\,s), when the defect density is high, many of the defects have a separation that is smaller than the imaging and machine learning spatial resolution limits, and the number of defects counted is lower than theoretically predicted.
At later times, the defect density exhibits power law decay with an exponent of $0.9$, in agreement with theory.

\section{\label{sec:exp}Experiment}

\begin{center}
\begin{figure}
\includegraphics[width=3.5in]{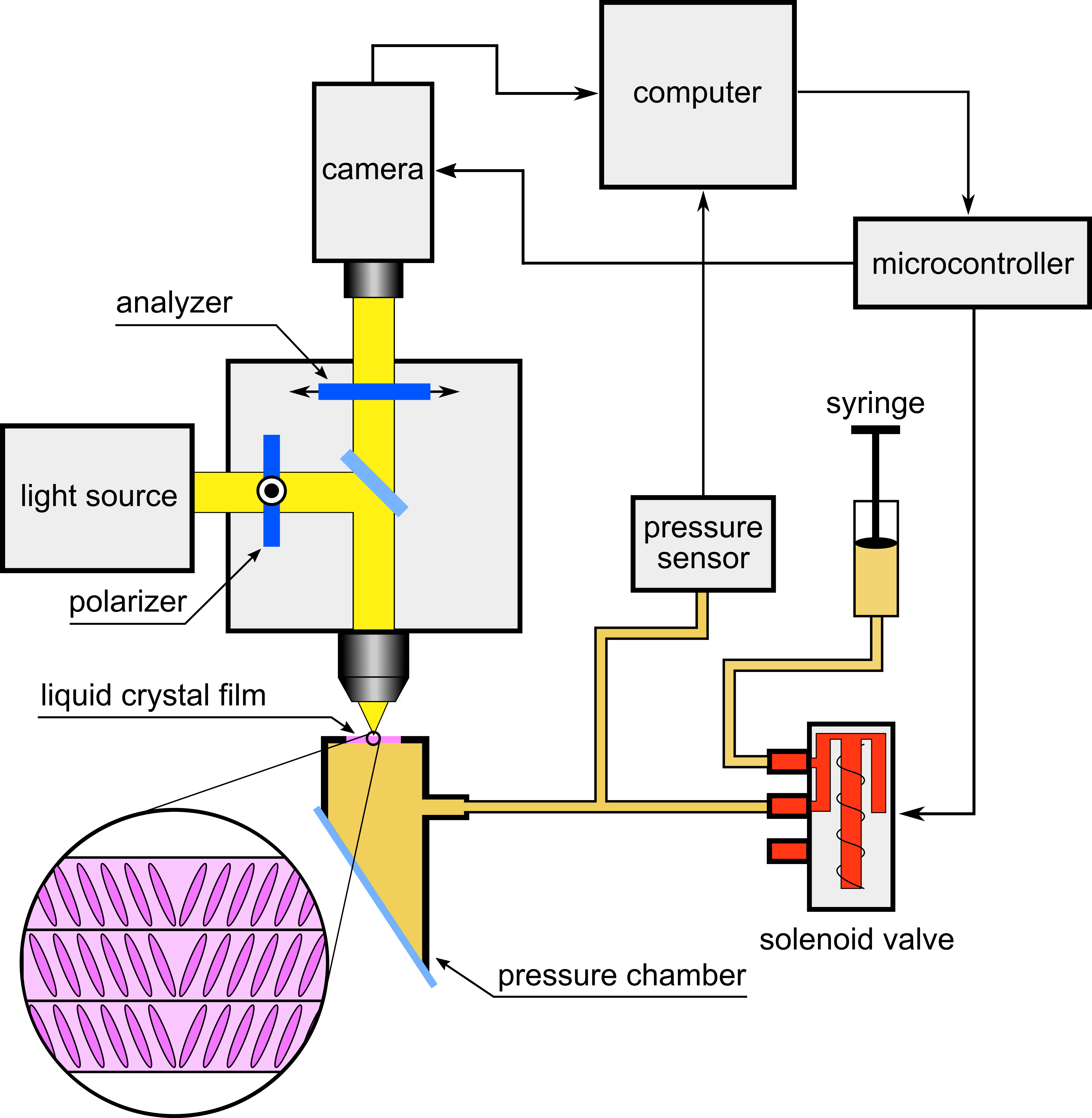}
\caption{\label{fig:experiment}Schematic of the film quenching experiment. A thin smectic-C film  drawn across a small, circular opening in a sealed chamber is temporarily deformed to a dome by increasing the chamber pressure. A sudden release of the excess pressure then causes the film to return rapidly to its original planar geometry, a mechanical quench that results in the spontaneous formation of topological defects in the director field (inset).
The defects are visualized using polarized reflected light microscopy and the coarsening of the defect texture recorded using a high-speed video camera.}
\end{figure}
\end{center}

In the quenching experiments, smectic-C films are drawn across a circular aperture in a glass cover slip set in the opening of an otherwise airtight chamber. Increasing the air pressure in the chamber causes the originally flat film to be distorted into a dome. When the pressure is suddenly released, the film collapses rapidly to being planar again, a mechanical quench that increases the hydrostatic pressure in the film, causing a short-lived transition to the smectic-A phase. The subsequent return to the smectic-C phase results in the spontaneous appearance of thousands of $2\pi$ disclinations, topological defects of unit strength (i.e., with winding numbers $\pm 1$), in the film. This initially dense array of defects then coarsens by the mutual annihilation of defects of opposite sign \cite{muzny_thesis_1994}. The experimental setup is shown in Fig.~\ref{fig:experiment} and described in further detail by Green \cite{green_liquid_2019}. The evolution of the defect texture was captured using a high-speed video camera (Phantom V$12.2$) with a spatial resolution of $1104\times{}800$ pixels and a bit depth of $16$, operating at $500\,$fps.

The liquid crystal material used in these experiments was PM$2$, a $50$:$50$ mixture by weight of SYNTHON ST$00552$ ($2$-($4$-$n$-hexyloxyphenyl)-$5$-$n$-octylpyrimidine) and SYNTHON ST$00557$ ($5$-$n$-decyl-$2$-($4$-$n$-octyloxyphenyl)pyrimidine) \cite{synthon}, with the phase sequence
SmC $52\unit{\celsius}$ SmA $68\unit{\celsius}$ N $72\unit{\celsius}$ Iso \cite{harth_episodes_2016}. Films $5\,$mm in diameter and  $20$--$30$ molecular layers ($60$--$90\,$nm) thick were drawn in the SmC phase at room temperature and the quenching experiments conducted at $35\unit{\celsius}$.

The $c$-director field in a ${\sim}0.6\,$mm$^2$ region of the film was visualized using polarized reflection microscopy. Under crossed polarizers, the film displays a schlieren texture, with each defect core surrounded by four alternating dark and light brushes. In order to be able to visualize the coarsening dynamics at high frame rates, the average intensity of the image was increased by decrossing the polarizers. This halves the number of brushes around each defect, transforming the cross-like brush textures to bow ties \cite{link_thesis_1998}. At the earliest times that defects can be observed following a typical quench, the average separation between defects is typically around $20\,\mu$m. Defects of opposite sign exert long-range $1/r$ attractive elastic forces on one another, while those of the same sign repel, interacting like infinite lines of electrical charge \cite{muzny_thesis_1994}. The defects also exhibit Brownian motion, diffusing laterally, in the plane of the film \cite{muzny_direct_1992}. Over time, neighboring $+1$ and $-1$ defects approach each other and mutually annihilate, with most of the defects disappearing within a few seconds of the quench. A typical coarsening sequence is shown in Fig.~\ref{fig:sequence}.

\begin{center}
\begin{figure}
\includegraphics[width=3.5in]{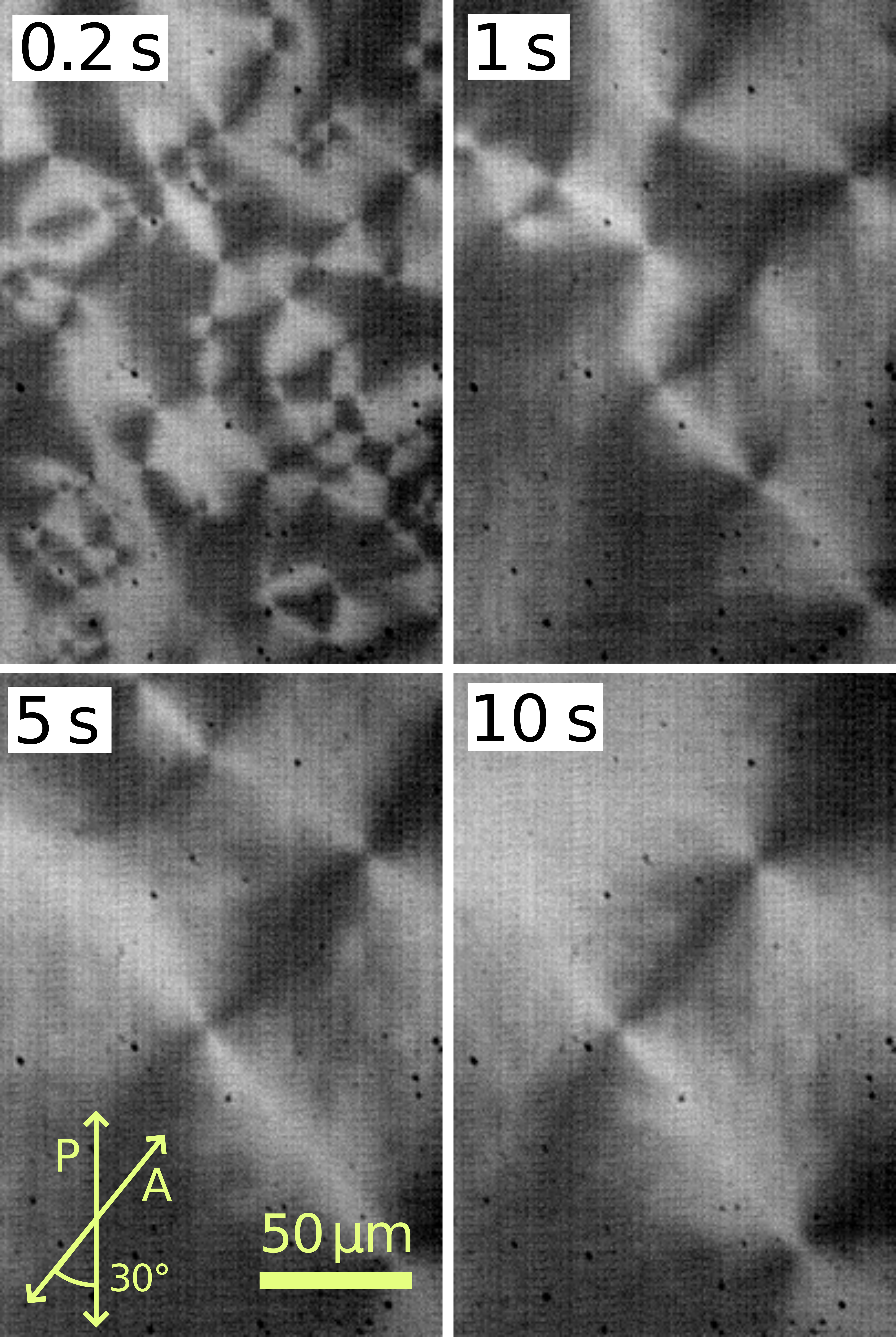}
\caption{\label{fig:sequence}Defect coarsening in a mechanically quenched SmC film. The topological defects are located at the points from which the pairs of light and dark brushes (resembling bow ties) emanate. Thousands of $+1$ and $-1$ defects are generated during a typical quench and these mutually annihilate over time. The polarizer and analyzer are decrossed by $60\unit{\degree}$. The intensities of these images, which show only a small part of the film, have been normalized as described in the text.}
\end{figure}
\end{center}

\section{\label{sec:yolov5}Topological Defect Detection}

A rigorous analysis of the dynamics of topological defects in smectic-C films requires identifying their topological signs and tracking their locations as a function of time. Conventional feature-detection algorithms that use intensity thresholding or edge-detection of object boundaries have been successfully used in many  investigations of soft materials to detect features with regular shapes and well defined boundaries, such as colloidal particles in suspension\cite{crocker_methods_1996} and islands and droplets on smectic films \cite{qi_mutual_2014,qi_incompressible_2016,qi_active_2017,hedlund_detection_2022}.
Detecting topological defects in liquid crystals is, however, a much more challenging task because the defects are identified principally by the diffuse schlieren textures surrounding them, which  are typically irregular, have no well-defined boundaries, and vary in appearance. Quenching a film generates thousands of defects, resulting in a complex schlieren texture with defect core separations of as little as a few microns in our experiment. Accurate analysis of such textures is not feasible using our previous methods, in which we detected and tracked well-separated defects in equilibrium films by cross-correlating the experimental images with synthetically generated defect templates to determine their locations and brush orientations \cite{silvestre_modeling_2009}.

We recently demonstrated the utility of modern deep learning networks for defect detection, using the  YOLOv2 network \cite{redmon_yolo9000_2016} trained on a large set of computed images of defect textures generated by Monte-Carlo simulations of the $2$D XY model \cite{minor_end--end_2020}. Although these training images were modified to emulate the experimental images more closely, nevertheless, several features present only in the experimental images regularly caused the trained model to predict false positives during analysis. For example, because dark speckles of the kind seen in many of the experimental images were not present in the training images, the network was not able to recognize these as being artifacts rather than defects. In addition, the black mask of the microscope field stop was not considered in the computed training images, resulting in false positives along the edges of the field of view.

In the present analysis, we used YOLOv$5$, a deep learning network designed for fast object detection \cite{jocher_ultralyticsyolov5_2021}, that was trained on real experimental images, allowing us to achieve a much higher detection accuracy than before. YOLOv$5$ is (at this time) the newest iteration of YOLO, a lineage of neural networks that perform both bounding box predictions (object localization) and classification tasks for every object in the image, in a single instantiation of the network. Technical details of the neural network are given in Appendix~\ref{app:architecture}.

\subsection{\label{sec:data}Image processing}

Images captured during nine different quenching experiments were analyzed. To compensate for variations in the brightness and contrast of these data sets, all of the images were normalized to have the same average intensity and dynamic range. This ensured that all of the images input to the neural network had similar statistical properties, making it easier for the correct weights to be developed in training. This reduced the number of false positives and significantly reduced the number of training epochs required.  A typical example of image normalization is shown in Fig.~\ref{fig:normalization}.

\begin{center}
\begin{figure}
\includegraphics[width=3.5in]{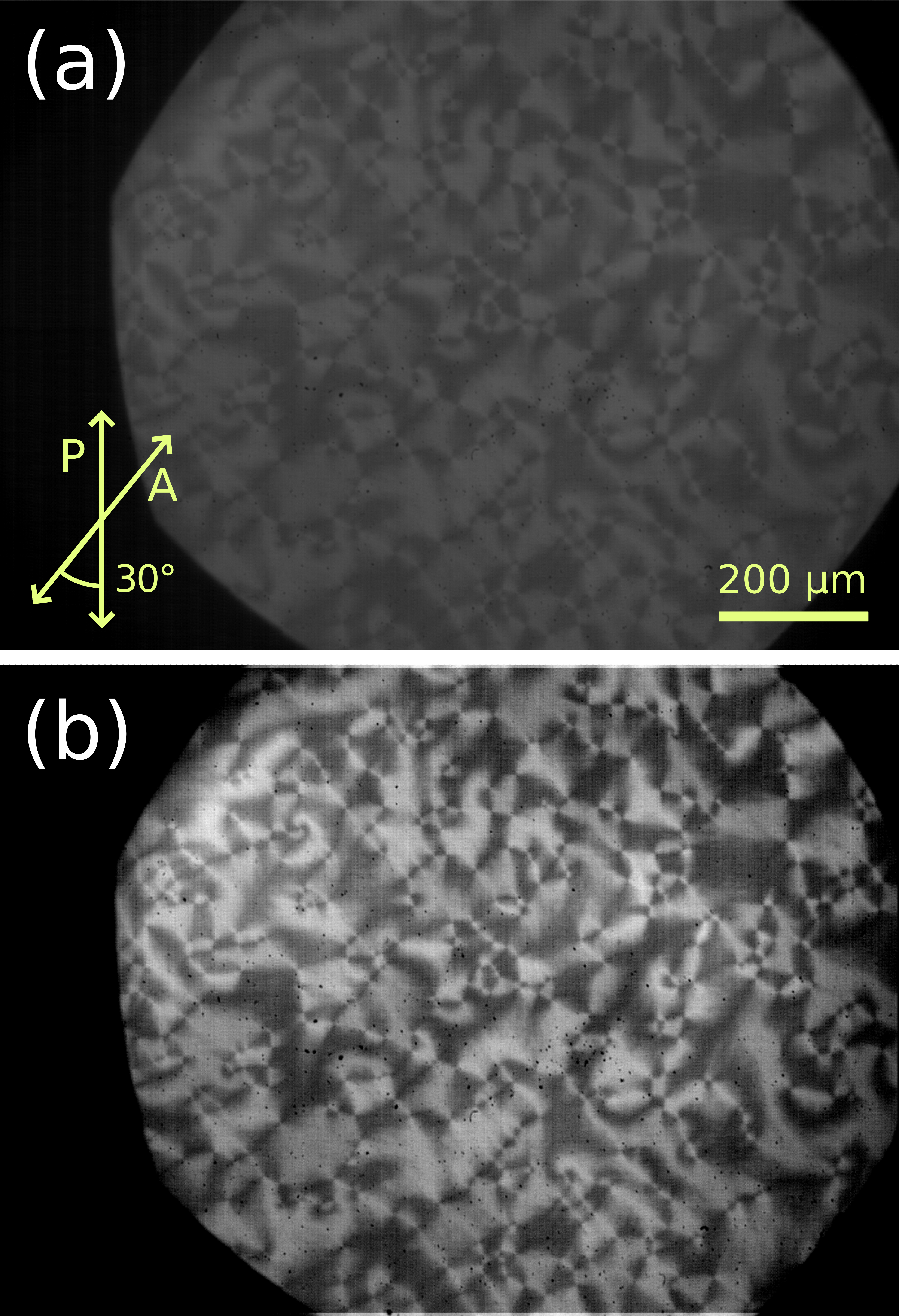}
\caption{\label{fig:normalization}Normalization of experimental images. (a) Raw experimental image ($t=0.4\,$s). (b) Normalized image. The raw images extracted from the quenching videos were normalized to have the same mean image intensity and dynamic range. This procedure ensured that the statistics of all of the experimental images from different quenching events would match those of the training set.}
\end{figure}
\end{center}

\subsection{\label{sec:train}Model Training}

The neural network was trained on experimental data, using a set of $141$ images chosen at random from different film quenching experiments and divided into training and testing sets in the ratio $80$:$20$. Gradient-descent calculations were carried out using the training set and the model performance was logged every cycle using the testing set. The locations of the defects in the training data (the `ground truth' locations) were determined manually.

Training was carried out using Google's cloud research computing service, the Google Colaboratory, on an NVIDIA Tesla T$4$ GPU with $16\,$GB of memory \cite{google_colaboratory}. Four YOLOv$5$ models of different sizes (listed in Appendix~\ref{app:architecture}) were trained for $500$ epochs each on the same training set, using a batch size of $16$. Model checkpoints (weights) were stored every epoch and the checkpoint yielding the highest mean average precision (mAP) on the testing set was selected. Choosing the optimal checkpoint of the neural network in this manner helped to prevent over-fitting, which is generally a concern when the training data sets have fewer than $500$ images, as in our case.

YOLOv$5$ uses several data augmentation techniques that further reduce the possibility of over-fitting by changing the appearance of the training images in order to increase artificially the amount of training data. These modifications include carrying out vertical and horizontal flips, cropping, rotation, and a new method introduced with YOLOv$5$ called mosaic augmentation that meshes sections of multiple images together. In addition, the image quality may be altered intentionally using randomized exposure, saturation, or blurring \cite{jocher_ultralyticsyolov5_2021}.

Square bounding boxes were used to define the defect core locations. While keeping the bounding boxes small has the benefit of increasing the precision of the detections and enabling defect detection even when the defects are close together, this also reduces the number of pixels associated with each defect, making training more difficult. The smallest bounding box size that our network could train on reliably was found to be $11\times{11}$ pixels.

\subsection{\label{sec:valid}Neural Network Performance}

The performance of the four trained neural network models was evaluated using a control set of $48$ normalized test images which were not included in the training data and hence had never been `seen' before by the network. The detection results were compared to manually obtained `ground-truth' locations using YOLOv$5$'s built in performance metrics. Of particular significance are the mean Average Precision (mAP) values and the precision recall (F$1$) scores, since these metrics contain information about both false negatives and false positives \cite{salton_introduction_1983}. The model with the highest recorded mAP score was the smallest model, YOLOv$5$s, which achieved an mAP of $0.970$ and a peak F$1$ score of $0.96$. The metrics computed for the various models are summarized in Table~\ref{tab:table} in Appendix~\ref{app:PR_and_F1}.

Although using image cross-correlation to identify defects yielded accurate results in experiments where the density of defects in the film was low \cite{silvestre_modeling_2009}, defect detection using this method becomes intractable in films with high defect densities, as is the case at early times in the quenching experiments described here. Cross-correlation is also sensitive to the presence of artifacts in the image, such as the liquid crystal deposits on the film chamber window visible as black speckles in most of our experimental images of quenched films. The resulting false predictions cause this method to have a low mAP, typically around $0.5$ (see Table~\ref{tab:table} in Appendix~\ref{app:PR_and_F1}). Deep learning networks, in contrast, can be trained to ignore such artifacts. The machine learning method developed by Minor et~al.\ using YOLOv2 trained on synthetic images to analyze the quenching experiments occasionally mis-identified the black speckles in the experimental images as topological defects \cite{minor_end--end_2020}, and the highest mAP achieved with this approach was around $0.8$. In the present study, the YOLOv$5$ network trained on real images that included speckles and the edges of the field stop obtained significantly better mAP scores.

\section{\label{sec:track}Defect Tracking and Sign Classification}

\subsection{\label{sec:link}Trajectory Linking}

Once the defect locations had been determined in each frame, linked trajectories were constructed using TrackPy \cite{trackpy}, a Python implementation of an algorithm originally developed for tracking colloidal particles \cite{crocker_methods_1996}. The `memory' feature of this program enables complete trajectories to be constructed even when there are occasional missed detections.

Visual comparisons of the computed defect trajectories following several different quenching events with manually determined defect locations demonstrated broad agreement, providing validation of the predictions of the neural network. A typical example of computed defect trajectories is shown in Fig.~\ref{fig:tracking}.

Two key assumptions about the defect dynamics were made in linking the defect locations to form trajectories. First, it was assumed that all of the topological defects present in the field of view were created at the time of the quench. Second, we assumed that defects of opposite topological sign annihilate and disappear in a pairwise manner. Defects that entered or left the frame during the coarsening experiment were exceptions that required special treatment when constructing their trajectories.

\begin{center}
\begin{figure}
\includegraphics[width=3.5in]{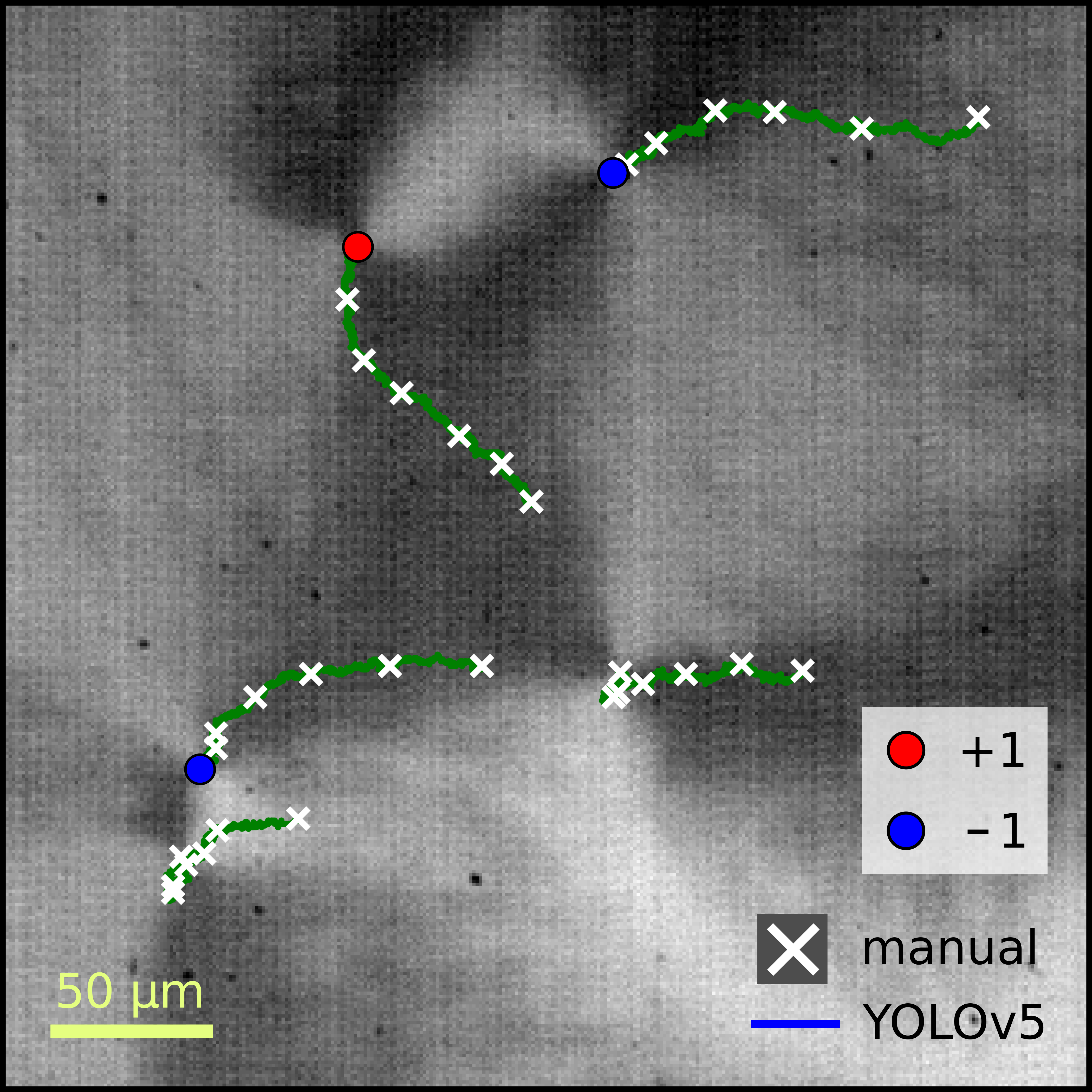}
\caption{\label{fig:tracking}Defect trajectories in a quenched film as determined by the trained neural network (green tracks) compared with defect locations identified manually (white crosses) every $500$ frames (at $1$ second intervals).
The  trajectories, which were determined over the course of $3000$ frames (an elapsed time of $6$ seconds), are superimposed here on the final image in the video sequence, with the surviving $+1$ and $-1$ defects shown respectively in red and blue. The dark speckles are liquid crystal droplets deposited on the chamber window by previously ruptured films. The neural network was trained to ignore these artifacts. The polarizer/analyzer settings are as in Fig.~\ref{fig:normalization}.}
\end{figure}
\end{center}

\subsection{\label{sec:orientation}Defect Sign Classification from Brush Orientation Dynamics}

When SmC films at equilibrium are observed in real time in the polarized light microscope, the signs of any topological defects may be readily determined by judiciously varying the decrossing angle of the polarizers and rotating the film. This procedure can not, however, be followed in the quenching experiments, where most of the defects disappear less than a second after the quench occurs.
The $+1$ and $-1$ defects are generally very similar in appearance and distinguishing them in static images alone is difficult. Near their cores, both kinds of defect resemble bow ties when the polarizers are decrossed but their orientations and the schlieren textures around them are highly variable, depending in a non-trivial way on their locations relative to other defects in the film. We have nevertheless been able to solve this fundamental classification problem by using their characteristic orientational dynamics to discriminate between the $+1$ and $-1$ defects.

First, the orientations of all of the defects, by which we mean the orientations of the bright brushes in the schlieren texture around the defects, were determined in every video frame. This was achieved by the previously mentioned technique of cross-correlating the region around every defect core with small, synthetic defect templates generated at different angles. When the computed defect orientations are plotted as a function of time, as in the example of Fig.~\ref{fig:orient}, it is immediately apparent that the defects fall into two classes. The brush orientations of the first class of defect show little variation over time, with only small azimuthal fluctuations of less than $10\unit{\degree}$.
The mean orientation of these defects was found, in all of the quenched films, to be ${\sim}135\unit{\degree}$.
The brush orientations of the second class of defects, in contrast, are not confined to a particular azimuth and vary substantially over time.

Since $+1$ defects have full axial ($C_\infty$) symmetry, their brush orientation is relatively insensitive to thermal orientation fluctuations of the $c$-director.
Defects of strength $-1$, on the other hand, are characterized by alternating bend and splay distortions of the surrounding $c$-director field and have only two-fold ($C_2$) symmetry. A consequence of this anisotropy is that the appearance of the $-1$ defects is effectively more sensitive to thermal fluctuations and to their environment, responding readily to reorientations of the surrounding $c$-director field caused, for example, by distant defect annihilations or resulting from spatial rearrangements associated with the approach of other defects in the film.
The orientational mobility of the brushes around $-1$ defects was previously reported by Wachs \cite{wachs_dynamics_2014}, who observed that in thin films, an isolated $-1$ defect typically reorients shortly before annihilating with a $+1$ defect, maintaining the continuity of the director field along the line connecting the two defects. The appearance of the $+1$ defect, in contrast, seems to be relatively unaffected by the impending annihilation. This phenomenon has recently also been observed by  Missaoui et~al.\ \cite{missaoui_annihilation_2020} in thick SmC films and modelled  theoretically  by Tang and Selinger \cite{tang_annihilation_2020}.
We conclude, therefore, that the first observed class of defects in our experiments has topological charge $+1$ and the second class topological charge $-1$.

\begin{center}
\begin{figure}
\includegraphics[width=3.5in]{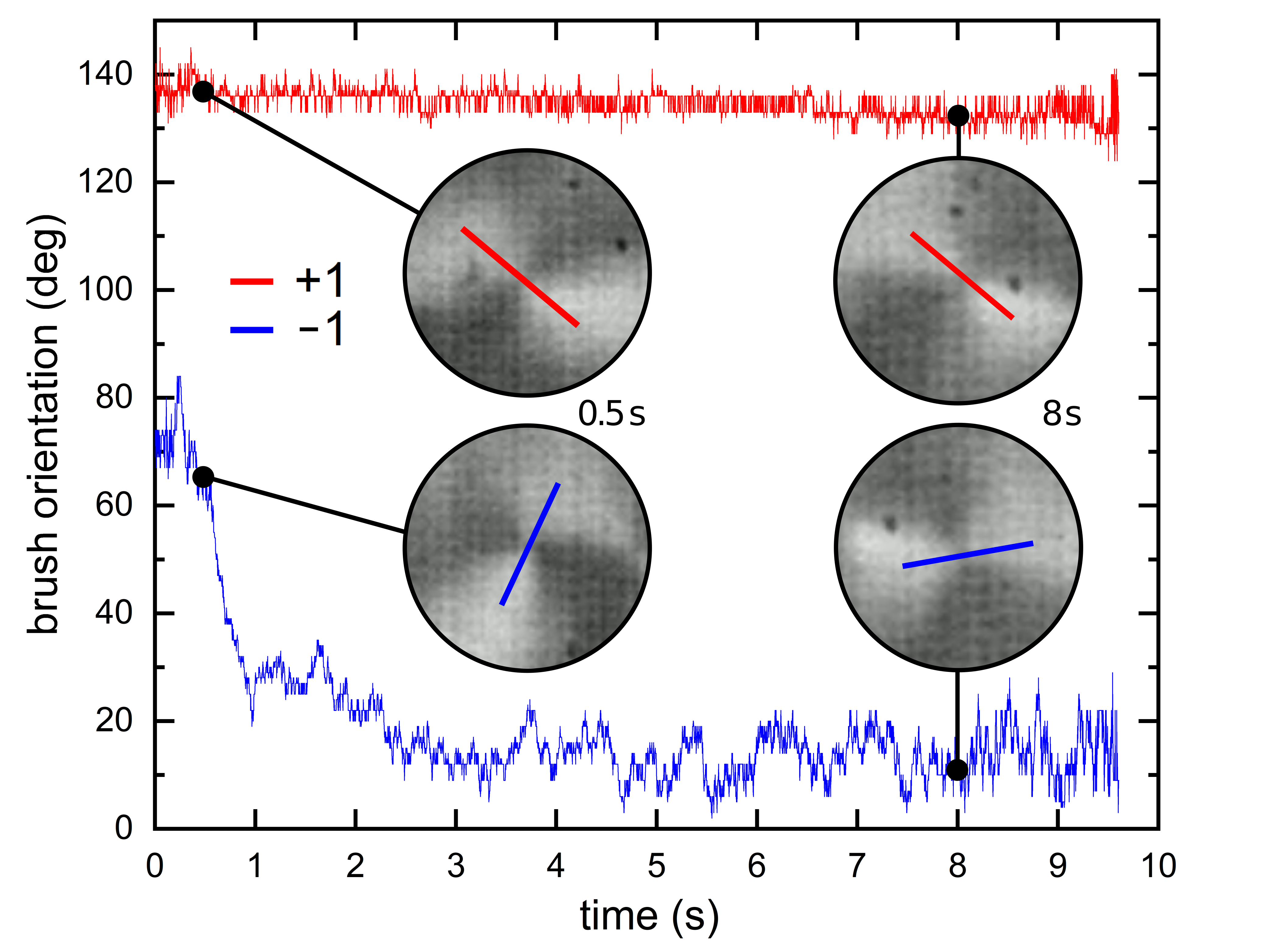}
\caption{\label{fig:orient}Bright brush orientation as a function of time for a typical pair of annihilating defects.
Quenching occurs at $t=0$  and the pair annihilates at $t=9.6\,$s. Over the lifetime of the pair, the brushes around the $+1$ defect (red trace) fluctuate about an azimuth of $135\unit{\degree}$  but do not change their average orientation significantly. The $-1$ defect (blue trace), in contrast, is more orientationally mobile. Initially oriented at $75\unit{\degree}$ in this example, the brushes soon rotate to around $20\unit{\degree}$, where they remain until annihilation. The insets show snapshots of the defects shortly after the quench ($t=0.5\,$s) and shortly before annihilation ($t=8\,$s). The variations in brush orientation over time are evaluated by a binary classification network in order to determine the topological signs of all of the defects.}
\end{figure}
\end{center}

Second, determination of defect sign on the basis of the brush orientation dynamics was achieved using a custom binary classification network comprising a fully connected model with one hidden layer. The hidden layer adds the complexity necessary to account for nonlinearities in the relationship between brush orientation dynamics and defect strength. The orientational data were reduced to three input features for each defect: the mean brush orientation, the standard deviation of the orientation, and the defect lifetime. The model determined the probability of each defect having strength $+1$ or strength $-1$. Defects with readily identifiable strengths were selected manually to generate training data for the model. $74$ such defects were labelled in total, $60$ for training and $14$ for testing. The model was trained for $1000$ epochs.

\section{Results}

\subsection{\label{sec:detection}Detection and Classification of Defects}

As we have seen in Fig.~\ref{fig:tracking}, the defect locations determined by YOLOv$5$ are in very good agreement with those determined manually.  Starting at early times, soon after the quench ($t<1\,$s), the network even identifies defects in high-density regions that are missed by visual inspection. The numbers of $+1$ and $-1$ defects identified in each video frame were found to be roughly equal, as expected. Small deviations from parity are inevitable since only a finite region of the film is imaged, causing some of the partner defects to be located outside the field of view at any given time. Visual inspection confirmed that the number of false positives in any image was typically very small, even near the field stop defining the edges of the field of view or in the presence of artifacts in the image. An example of defects detected by the neural network and classified by topological sign is shown in Fig~\ref{fig:anno}.

\begin{center}
\begin{figure}
\includegraphics[width=3.5in]{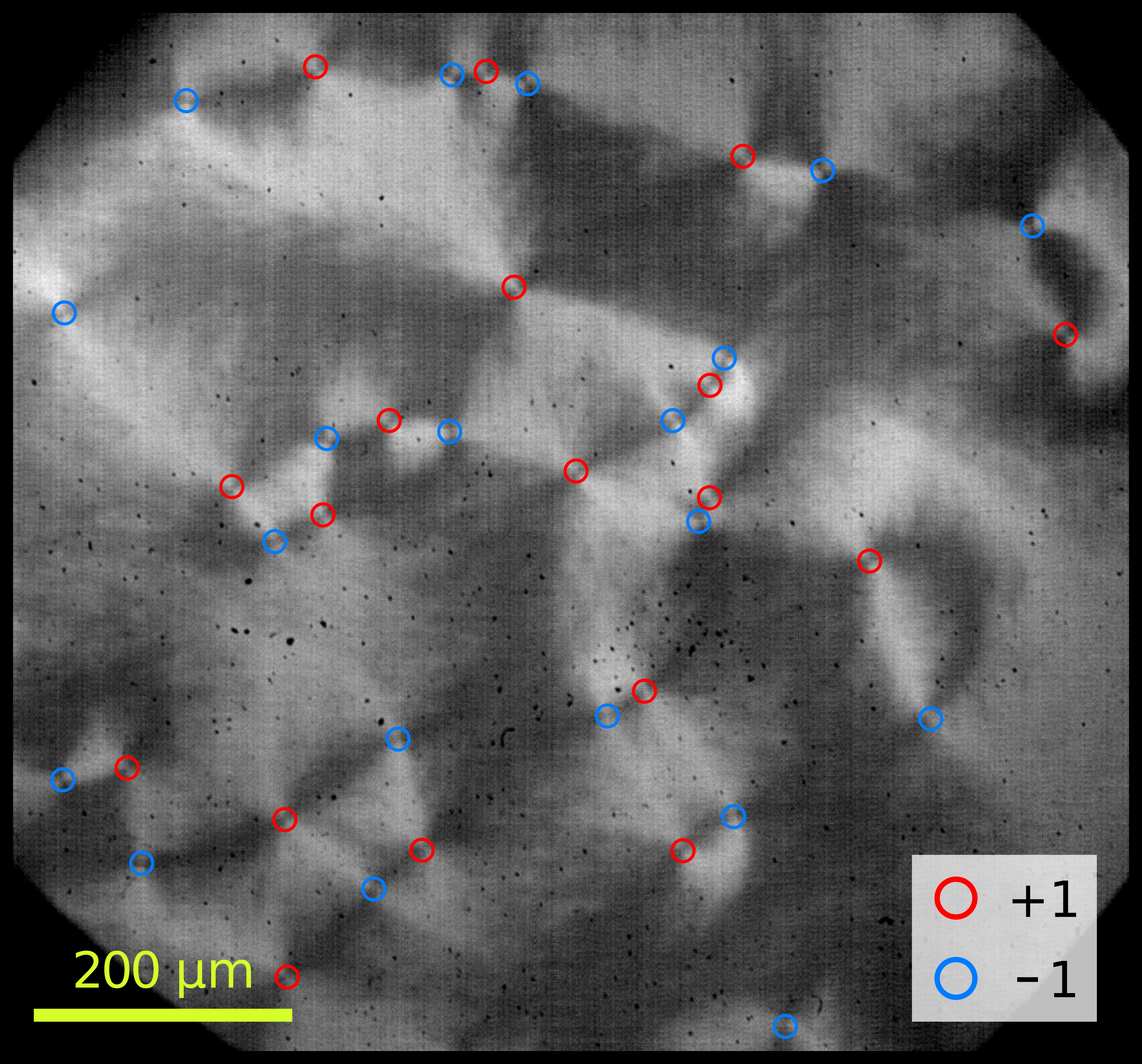}
\caption{\label{fig:anno}Defect locations predicted by the neural network and classified according to topological sign. The image shows a SmC film six seconds after quenching, with the defects color-coded according to the topological sign predicted by the custom binary classification network. The defect locations determined by the deep learning model were proved generally to be highly accurate.  In this example, every defect was detected: $18$ of strength $+1$ and $21$ of strength $-1$.
The film diameter is $5\,$mm, much larger than the image, the black borders here corresponding to the edges of the microscope field stop. The polarizer/analyzer settings are as in Fig.~\ref{fig:normalization}.}
\end{figure}
\end{center}

\subsection{\label{sec:coarsening}Coarsening Dynamics}

The decay of $N(t)$, the number of topological defects, as a function of time measured in nine different quenching experiments showed similar behavior. Defects could only be detected starting about $0.2\,$seconds after the quench was initiated, when the previously distorted film was flat and in focus again. In every experiment, $N$ was observed to decrease slowly at first and then more quickly, decaying algebraically at longer times, with a roughly constant exponent. A typical measurement of $N(t)$ is shown in Fig.~\ref{fig:N_vs_t}.

The overdamped, continuous XY model describing locally interacting, classical spins in $2$D predicts an inverse power-law relation, $N \propto t^{-1}$ \cite{chaikin_principles_2013}. It is readily apparent from the graph, however, that the observed decay occurs more slowly overall than predicted by this model and deviates significantly from a simple power law at early times. Yurke et~al.\ carried out computer simulations of the $2$D~XY model and described the observed coarsening behavior with $N \ln{N} \propto t^{-1}$, the logarithmic correction accounting for the effective drag on the defect cores \cite{yurke_coarsening_1993}. Jang et~al.\ also carried out numerical simulations including these  frictional forces, finding that $N$ varied as  $N \propto t^{-0.9}$ \cite{jang_annihilation_1995}, in agreement with the asymptotic scaling behavior predicted by the Yurke model.
As is evident from Fig.~\ref{fig:N_vs_t},  our experimental data are fit well  by the Yurke and Jang predictions at intermediate and long times ($t \agt 0.4$\,s).

\begin{center}
\begin{figure}
\includegraphics[width=3.5in]{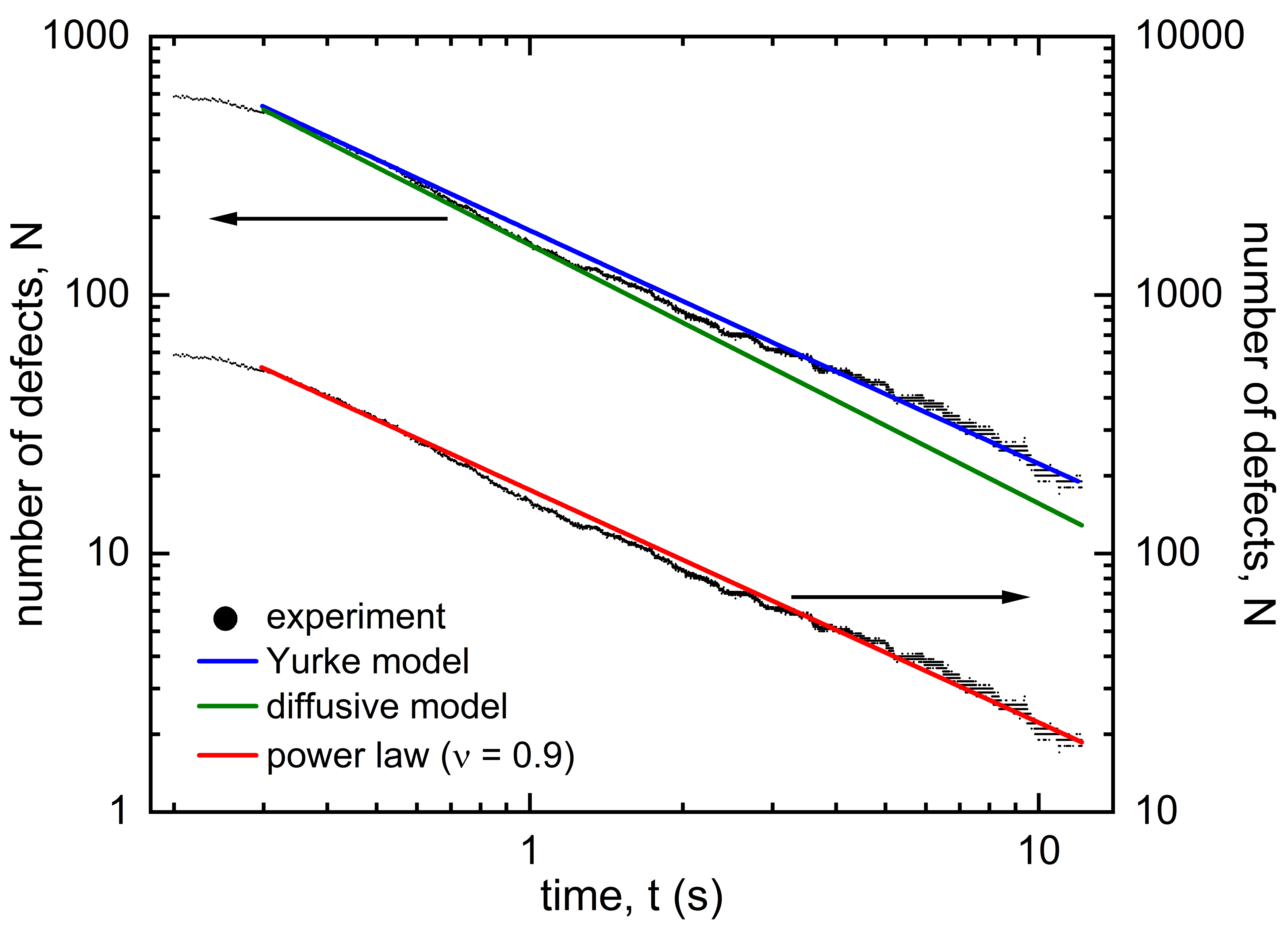}
\caption{\label{fig:N_vs_t}Defect number vs.\ time in a quenched smectic-C film. The large number of topological defects generated in a typical quenching experiment decreases over time by the mutual annihilation of $+1$, $-1$ defect pairs. Images of the film were analyzed starting when the film came back into focus, about $0.2$ seconds after the mechanical quench. A fit with the diffusive model (green), which describes algebraic decay of the defect number $N$ with an exponent of $1$,  predicts coarsening at long times that is faster than observed. The logarithmic correction term of Yurke's model (blue) results in a slower annihilation rate that closely approximates the experimental measurements. Power-law decay with an exponent of $0.9$, as suggested by the simulations of Jang et~al., is in accord with the Yurke model and matches the data similarly well (red). The experimental data is duplicated here for visual clarity.}
\end{figure}
\end{center}

At early times, between $0.2$ and $0.4$ seconds after the quench, the observed dynamics deviate significantly from  power-law behavior. Extrapolating the asymptotic slope to early times leads to the prediction that at $0.2$ seconds there should be around $800$ topological defects, about $200$ more defects than were detected by the neural network. This discrepancy is attributed to the undercounting of defects that are closer than about $10\,\mu$m ($11$ pixels), which is the resolution limit for detection by the trained network. This limit is also manifest in defect pair correlation functions computed from the images, which are identically zero for defect separations closer than $10\,\mu$m, as shown in Fig~\ref{fig:correlation} in Appendix~\ref{app:correlation}. Comparing the pair correlation curves derived from experimental data with those computed from  simulations, also plotted in Fig~\ref{fig:correlation}, supports the notion that the neural network is unable to discern topological defects that are very close together in the experimental images, resulting in undercounting at early times when the defect density is highest.

Finally, it has been suggested that there are circumstances in which the $2$D~XY model would be expected to exhibit exponential rather than power-law decay at early times (see, for example, \cite{toussaint1983,loft1987}), which would clearly result in a knee of the kind shown in the plot of Fig.~\ref{fig:N_vs_t}. However, the inherent uncertainty in our early-time data precludes any systematic consideration of this possibility.

\section{\label{sec:summary}Summary}

In summary, we have demonstrated a deep-learning approach that allows us to detect topological defects in thin smectic-C liquid crystal films with a high level of accuracy. We have also developed a rigorous method for classifying the topological signs of the defects, using the distinctive orientational dynamic behavior of the director fields around them. A binary classification network trained to perform this task gives predictions consistent with the known physical properties of such arrays of defects, for example that mutual annihilation only occurs between defects of opposite sign and that in any given region of the film, the numbers of $+1$ and $-1$ defects are expected to be approximately equal.

We compared our experimental observations of the defect coarsening dynamics in films with several theoretical models. At long times, the number of defects was observed to decrease more slowly than predicted by the purely diffusive XY model, showing instead the power-law behavior with an exponent of $0.9$ in agreement with the model of Yurke et~al.\  and the simulations of Jang et~al. The anomalously slow annihilation rate observed at early times is attributed to the undercounting of defects, an unavoidable consequence of the inability of the neural network to resolve defects with separations smaller than $10\,\mu$m.

The deep-learning method described here could be improved by designing custom networks explicitly for detecting topological defects, which could allow a reduction in the number of required layers in the neural network, making defect detection faster and more transparent. The kind of machine learning implemented here could readily be applied to other detection tasks in soft materials, such as tracking a variety of inclusions in smectic films, colloidal particles in suspension, and bacterial cells in fluid media, or could be used to analyze the evolution of bubbles in foam coarsening experiments.

\begin{acknowledgments}
The authors wish to acknowledge many useful discussions with Matt Glaser.
This work was supported by NASA Grants NAGNNX07AE48G and NNX-13AQ81G, and by the Soft Materials Research Center under NSF MRSEC Grants DMR 0820579 and DMR-1420736. R.A.C. was supported by a UROP research award from the University of Colorado Boulder.
\end{acknowledgments}

\appendix

\section{\label{app:architecture}The Neural Network}

\begin{center}
\begin{figure}
\includegraphics[width=3.5in]{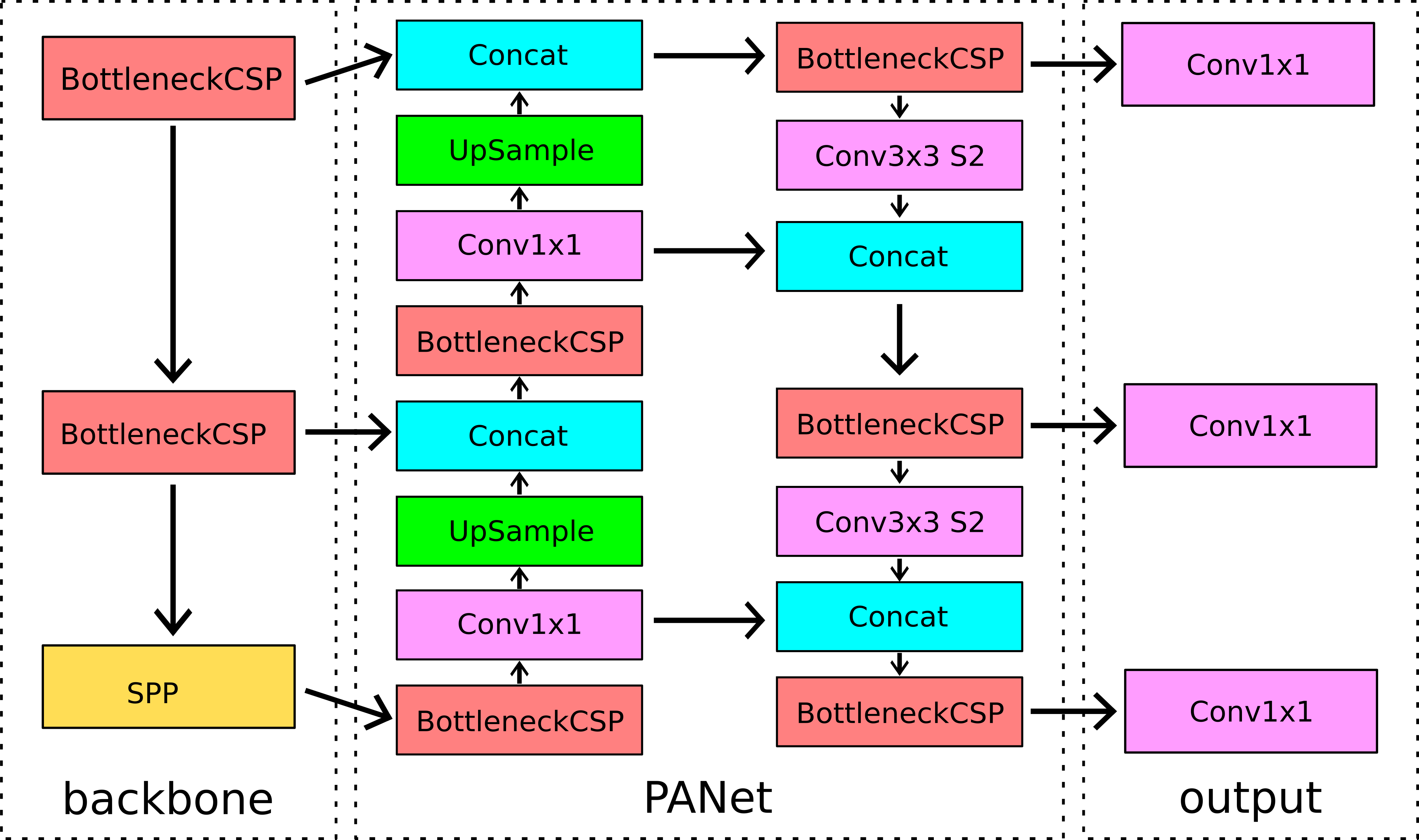}
\caption{\label{fig:architecture}Architecture of the YOLOv$5$ neural network. YOLOv$5$ comprises three parts, shown in the diagram: The model backbone is a Cross Stage Partial Network (CSPDarknet) for feature extraction. A Path Aggregation Network (PANet) combines features, which are then passed to the output layer for detection.}
\end{figure}
\end{center}

The architecture of YOLOv$5$, the neural network used in this study, is shown in Fig.~\ref{fig:architecture}. Unlike previous versions of YOLO that were developed in the Darknet framework, YOLOv$5$ is built using the PyTorch framework in Python. The backbone uses a Cross Stage Partial Network (CSPNet) to compress predicted features into fewer channels. A Spatial Pyramid Pooling Network (SPPNet) restructures the input to bypass fixed-size input constraints. The extracted features from the YOLOv$5$ backbone are then passed to a Path Aggregation Network (PANet) for feature fusion. Finally, the fused features are passed to the output layer (also called the YOLO layer), where the detection results are computed.

There are four popular YOLOv$5$ model sizes, with increasing numbers of layers: YOLOv$5$s, YOLOv$5$m, YOLOv$5$l, and YOLOv$5$x. We measured the training time,  mAP value, and peak F$1$ score of all four models. The training times, shown in  Table~\ref{tab:table}, ranged from $50$ minutes for the smallest model to $160$ minutes for the largest.

\begin{center}
\begin{figure}
\includegraphics[width=3.5in]{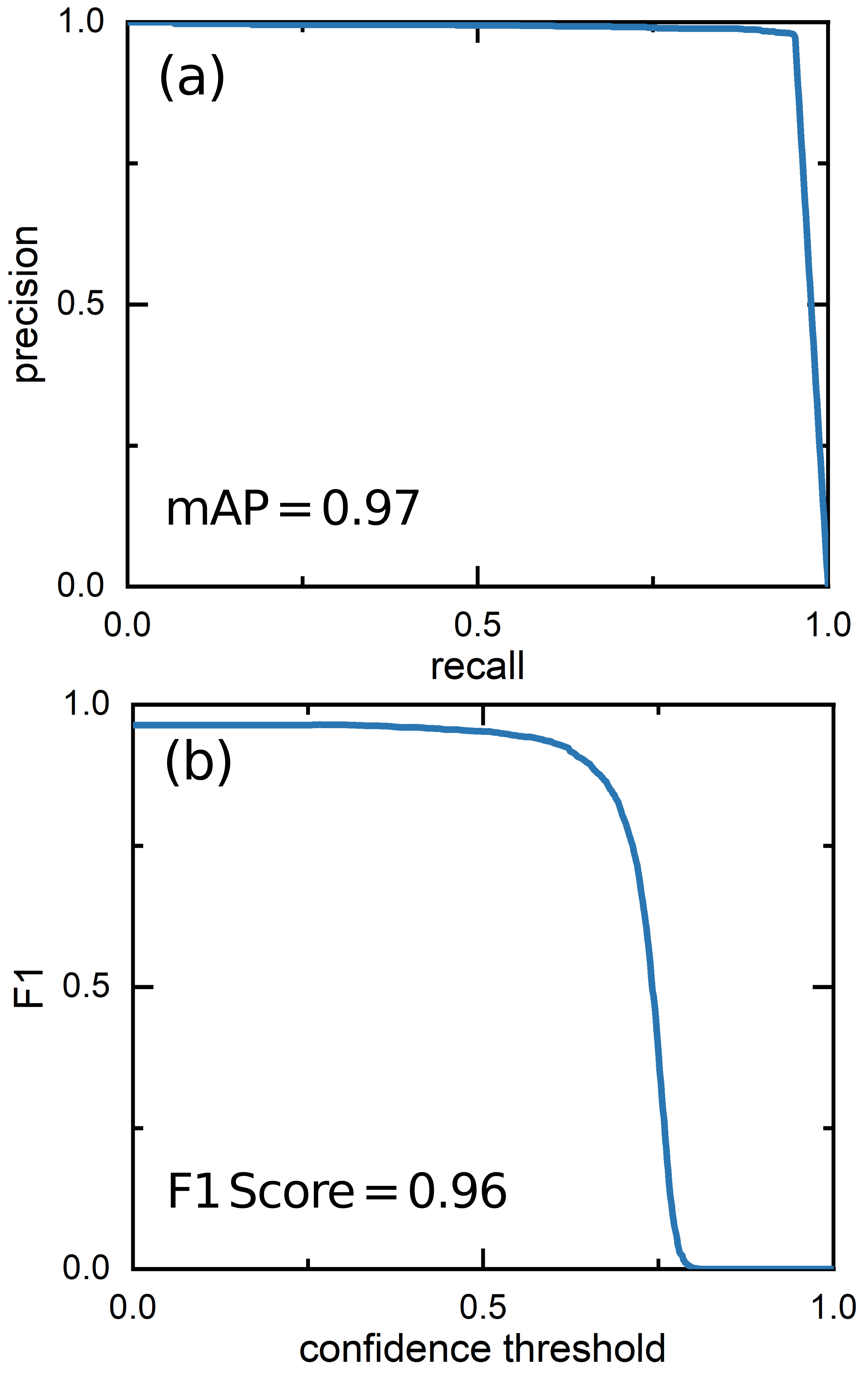}
\caption{\label{fig:metrics}Performance of the YOLOv$5$ model trained to detect topological defects.
(a) The precision-recall curve yields a mean Average Precision (mAP) of $0.97$, corresponding to a high degree of confidence in defect detection by the trained network. (b) The highest value on the F$1$ curve (the F$1$ score) was $0.96$. The best F$1$ score was obtained at a confidence threshold of $0.27$.}
\end{figure}
\end{center}

\section{\label{app:PR_and_F1}Precision-Recall Curve and F$1$ Plot}

In order to determine whether a detected result is a true or false positive, an intersection over union (IoU) of the detection bounding box with the ground truth bounding box is calculated. If the IoU ratio is above a pre-specified confidence threshold, the detection is classified as a true positive, while if it is below the threshold, it is deemed a false positive. The precision-recall (PR) curve is created by sampling results from a range of confidence threshold values, usually from $0.5$ to $0.95$. This is the range used, for example, by the Microsoft COCO data set, a large annotated collection of images with only a few sizeable objects per image \cite{fleet_microsoft_2014}. In our case, in contrast, there could be hundreds of defects per image with bounding box sizes as small as $11\times{}11$ pixels. For bounding boxes as small as these, small pixel-wise differences between manually created bounding boxes and those determined by YOLOv$5$ may impact the IoU dramatically. As a result, the likelihood of a true detection producing an IoU below $0.5$ is relatively high. To obtain a better measure of the network's results, we extended the threshold range from $0.0$ to $0.95$.

The mean average precision (mAP) is defined as the area under the precision-recall (PR) curve (Fig~\ref{fig:metrics}(a)), a plot of the ratio of true positives to all detections (precision) vs.\ the ratio of true positives to all ground truth values (recall). As seen in Table~\ref{tab:table}, the smaller model sizes were found to yield slightly better mAP scores in this study, with the smallest model, YOLOv$5$s, achieving the highest mAP score of $0.970$, marginally better than the score of $0.960$ achieved by YOLOv$5$x, the largest model we looked at. These results indicate that the additional layers present in the larger YOLOv$5$ models do not result in better defect detection.

The F$1$ score is the harmonic mean of the precision and recall values, calculated for a range of confidence thresholds. Since every threshold value has an associated F$1$ score, the threshold that yields the highest F$1$ score is routinely reported. As is evident from Fig~\ref{fig:metrics}(b), the highest F$1$ score (of $0.96$) occurs when no detections are ignored, demonstrating that our model is unlikely to detect false positives or make double detections (two detections of the same defect). The F$1$ score drops off rapidly above a confidence threshold of ${\sim}0.70$ and falls to zero before a threshold of $0.95$ can be reached.
This is a reflection of the sensitivity of the F$1$ score to small differences of a few pixels in bounding box locations and/or dimensions from the values manually determined in the control set. The highest F$1$ score for all four model sizes was $0.96$, as shown in Table~\ref{tab:table}. Also reported in the Table are the performance metrics for defect detection carried out both using template image cross-correlation and in the previous study using YOLOv2 \cite{minor_end--end_2020}.

\begin{table}
\caption{\label{tab:table}Neural network model training time and defect detection accuracy given by the mAP and F$1$ metrics.
Shown are the metrics for topological defect detection using different YOLOv$5$ models, as well as those achieved when cross-correlating the experimental images with synthetic defect templates and using YOLOv2 trained using synthetic images.}
\vspace{1 em}
\begin{ruledtabular}
\begin{tabular}{cccc}
Model                           & Training Time & mAP    & F$1$ Score   \\ \hline
Cross-Correlation               & --            & 0.498  & 0.68         \\
ForLL YOLOv2                    & 1.1 hours     & 0.818  & 0.81         \\
RealData11$\times$11 YOLOv$5$s    & 0.8 hours     & 0.970  & 0.96         \\
RealData11$\times$11 YOLOv$5$m    & 0.9 hours     & 0.967  & 0.96         \\
RealData11$\times$11 YOLOv$5$l    & 1.4 hours     & 0.963  & 0.96         \\
RealData11$\times$11 YOLOv$5$x    & 2.6 hours     & 0.960  & 0.96         \\
\end{tabular}
\end{ruledtabular}
\end{table}

\section{\label{app:correlation}Defect Pair Correlations}

The normalized pair correlation (radial distribution) function, $g(r)$, a measure of the average defect density as a function of distance from any given defect, has been computed from the experimental data as a function of time.
In these calculations, we used a radial bin size of $3\,\mu$m and the correlations were averaged over $10$ frames ($\Delta t = 0.02\,$s) in order to reduce the statistical noise.
We show in Fig~\ref{fig:correlation}, by way of example, the sign-agnostic defect pair correlation function at $t=0.2\,$s for the quench event analyzed in this paper.
This correlation function is identically zero below $10\,\mu$m because no topological defects closer than than this distance were identified by the neural network. This is an artifact of the detection process: defects with separations smaller than the bounding box size of $11\times{11}$ pixels (or about $10\,\mu$m) could not be detected.

A numerical simulation similar to \cite{jang_annihilation_1995} of the evolution of a large number (${\sim}15,000$) of diffusing, interacting topological charges initially generated with an average density similar to that observed in the film quenching experiments  was carried out. The pair correlation function derived from the simulated data, also shown in Fig~\ref{fig:correlation}, is seen to grow continuously from zero, as expected, saturating at long distances with a value $g(r)=1$. Comparing this to the experimental correlation function confirms that defects in the experimental images with separations below $10\,\mu$m are not detected by the neural network, leading to undercounting at early times following the quenching event.

\begin{center}
\begin{figure}
\includegraphics[width=3.5in]{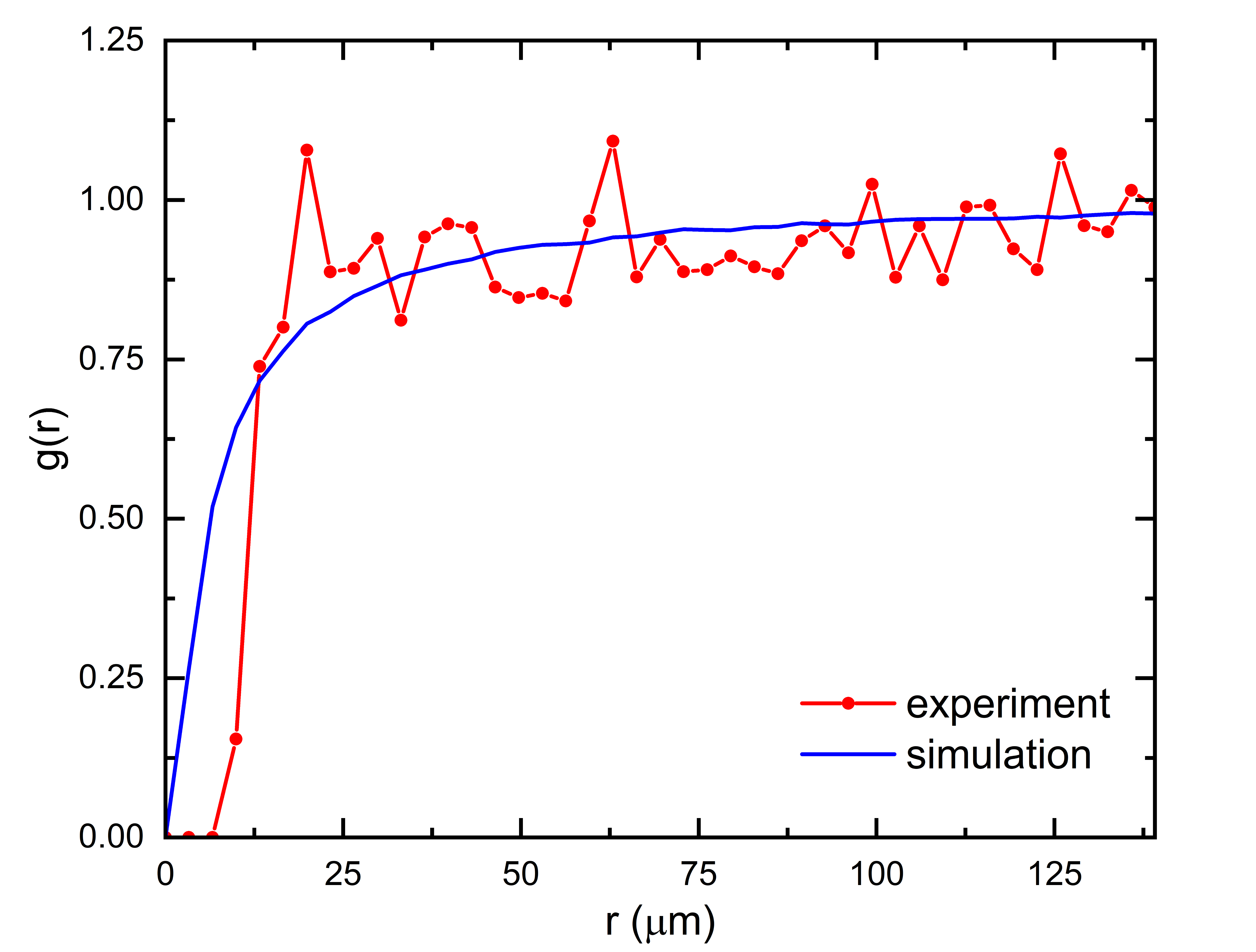}
\caption{\label{fig:correlation}Sign-agnostic topological defect pair correlation functions computed from experimental and simulation data shortly after a quench ($t=0.2\,$s). The correlation function derived from the simulation (blue) grows continuously from zero, while the experimental curve (red) has a sharp cutoff at $10\,\mu$m, the resolution limit for defect detection by the neural network.}
\end{figure}
\end{center}

\bibliography{defects}

\end{document}